\documentclass{ws-procs9x6}
\def\etal{{\sl et al.}}
\def\etc{etc.}
\let\ensm=\ensuremath
\newcommand{\gam}{\ensm{\gamma}}

\def\Pp{{\rm p}}

\def\piz               {\ensm{\pi^0}}

\newcommand{\be}{\begin{equation}}
\newcommand{\ee}{\end{equation}}
\newcommand{\rar}{\rightarrow}
\newcommand{\pdup}{p_\uparrow}
\newcommand{\nd}{\noindent}
\newcommand{\ppdup}{$\Pp + \Pp_{\uparrow} \rar \piz + X$}

\newcommand{\pdupp}{$\Pp_{\uparrow} + \Pp \rar \piz + X$}
\newcommand{\pimpr}{$\pi^- + \pdup \rar \pi^0 + X$}
\newcommand{\PR}[1]{{\it Phys.\ Rev.}\ {\bf #1}}
\def\gev              {\ensm{\,{\rm GeV}}}

\def\gevp             {\ensm{\,{\rm GeV}/c}}
\def\mevm             {\ensm{\,{\rm MeV}/c^2}}

\newcommand{\PRL}[1]{{\it Phys.\ Rev.\ Lett.}\ {\bf #1}}
\newcommand{\PhL}[1]{{\it Phys.\ Lett.}\ {\bf #1}}
\newcommand{\SNP}[1]{{\it Phys.\ of Atom.\ Nucl.}\ {\bf #1}}
\newcommand{\Yaf}[1]{{\it Yad.\ Fizika}\ {\bf #1}}
\def\xf{x_{\mathrm F}}

\begin{document}

\title{Single spin asymmetry measurements for $\pi^0$ inclusive
productions in \ppdup ~and \pimpr ~reactions ~at
70 and 40 GeV respectively. \footnote{\uppercase{T}his work is
supported by \uppercase{R}ussian \uppercase{F}oundation for
\uppercase{B}asic \uppercase{R}esearch, grant 03-02-16919}}

\author{A.M.~Davidenko, A.A.~Derevschikov, V.N.~Grishin,
V.Yu.~Khodyrev, Yu.A.~Matulenko, Yu.M.~Melnick, A.P.~Meschanin,
V.V.~Mochalov, L.V.~Nogach, S.B.~Nurushev\footnote{\uppercase{C}orresponding author,
E-mail:nurushev@mx.ihep.su}, P.A.~Semenov, A.F.~Prudkoglyad, K.E.~Shestermanov, L.F.~Soloviev,
 A.N.~Vasiliev, A.E.~Yakutin}

\address{Institute for High Energy Physics, Protvino,  Russia}
\author{ N.S.~Borisov, V.N.~Matafonov, A.B.~Neganov, 
Y.A.~Plis, Y.A.~Usov, A.N.~Fedorov}
\address{Joint Institute for Nuclear Physics, Dubna, Russia}
\author{ A.A.~Lukhanin }
\address{Kharkov Physical Technical Institute, Kharkov, Ukraine}
\author{ M.G.~Ryskin}
\address{Petersburg Nuclear Physics Institute, St. Petersburg,  Russia}

\maketitle

\abstracts{The inclusive $\pi^0$ asymmetries were measured in reactions 
$p+p\uparrow \rightarrow \pi^0+X$ and $\pi^-+p\uparrow \rightarrow \pi^0+X$ 
at 70 and 40 GeV/c respectively. The measurements were made at the central 
region (for the first reaction) and asymmetry is compatible with
zero in the entire measured $p_T$ region. For the second reaction the
asymmetry is zero for  small $x_F$ region ($-0.4<x_F<-0.1,~ 0.5<p_T(GeV/c)
<1.5$) and increases with growth of $\mid x_F\mid$. Averaged over the
interval $-0.8<x_F<-0.4,~ 1<p_T(GeV/c)<2$ the asymmetry was $-(13.8\pm
3.8)\%$.} 

Two experiments were performed at U70 GeV accelerator of IHEP aiming to 
measure the single spin asymmetry (SSA) in inclusive $\pi^0$ productions.
In one case we studied the reaction $pp_{\uparrow}\rightarrow \pi^0X$ at 70 GeV/c. The interest to such a study was stimulated by two experimental data
relevant to above reaction, namely, \cite{Dick} at 24 GeV/c and \cite{E704} at 200 GeV/c. 
The proton beam of 70 GeV/c momentum was  extracted from U70 by the thin
Si crystal bent under 80 mrad \cite{Aseev}.

\begin{figure}[t]
\centering
\includegraphics[width=\textwidth,height=4.5cm]
{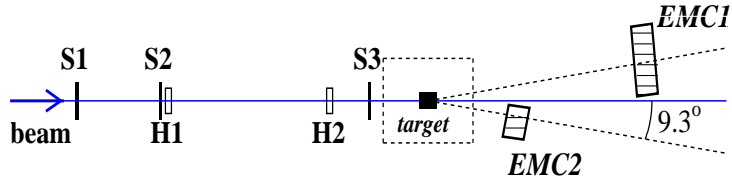}
\vspace*{-1.2cm}
\caption{ Experimental Set-up PROZA-M; S1-S3 -- Scintillation counters,
H1-H2 -- two-coordinate hodoscopes, EMC -- electromagnetic calorimeter,
{\it target} - polarized target.
\label{setup}}
\end{figure}

The beam parameters measured by counters S1-S3 and hodoscopes H1,H2 (see Fig.1) were: intensity 
$\approx 1\cdot 10^6$ p/cycle, 
$\sigma(x)=4$~mm, 
$\sigma(x ^{\prime})=2$~mrad, 
$\sigma(y)=3$~mm, $\sigma(y^{\prime})=1$~mrad.
The frozen spin polarized target (PT) was made of the propandiole, 
had the length of 200 mm, 
diameter of 20 mm. The average polarization was $\approx 80\%$.

Two electromagnetic calorimeters EMC1(480 lead glass counters(lgc)) and EMC2 (144 lgc) were used. They were positioned under $\pi^0$ emission angle
 $90^0$ in c.m.s. at 6.9 m and 2.8 m from PT correspondingly. Both calorimeters were composed from the same lead glass cell
of size 3.8x3.8x45 $cm^3$.
Calorimeters were calibrated by the electron beam of 26.6 GeV.
The energy resolution achieved was $\frac{\sigma (E)}{E}\approx 2.5\%$. 
The rate of data taking for pp-interactions was 350 events and during 10 days running $2\cdot 10^7$ events  were accumulated, while 
for $\pi^-p$ it  was around 300 events per
cycle and  $10^8$ events were accumulated for 30 days.

 The final results on the analyzing power for reaction \ppdup ~ at 70 GeV are
presented in Fig.2a (open square). It is evident that the SSA is compatible with zero.
 The study of the reaction \pimpr ~is similar to the previous one. The beam parameters were: intensity $1\cdot 10^6$ p/cycle, momentum of 40 GeV/c, the round beam size of 3.5 ~mm. In this case  we use only one EMC with 720 lgc exactly of the same type as before. 
The calorimeter was positioned  at
distance of 2.3 m from PT and angle was varied . The more details of the 
second experiment may be found in paper \cite{nuclphys40}. 
The results are presented in the Fig.2b (open square). If we 
average SSA over the interval $-0.8<x_F<-0.4$ we got 
$A_N(\pi^0)=-(13.8\pm 3.8)\%$, that is, an essential spin effect. But for the
interval $-0.4<x_F<-0.1$ $A_N(\pi^0)\approx 0$. 

    The general overview of the theoretical interpretations of our data at 40 and 70 GeV/c leads to 
the following understandings. Most of the theoretical models based on pQCD  are not applicable to our energy and transverse 
momentum ranges.  There are several phenomenological models (orbiting quarks, Sivers and Collins mechanisms, high twist effects, \etc )  explaining some aspects of data but not all of them. In this situation we applied the model proposed recently in paper \cite{rysk} to the present data.
. The slope parameter, needed in this model for calculation, was taken 
either from our experimental data or from the published data \cite{don}. The SSA results for reaction 
\ppdup ~at 70 GeV/c in CR \cite{nuclphys70} are presented in Fig.2a. The sign
of SSA as well as 
of $x_F$ were changed in order to compare to the E704 data presented in
the same Fig.2a. The dotted 
line is a prediction of the CMSM for 70 GeV/c and the solid line is for the
E704 data (see details in 
\cite{rysk}). There is fairly well consistency between experimental data
and the model prediction.

\begin{figure}
\begin{tabular}{cp{0.5cm}c}
\includegraphics[width=0.45\textwidth]
{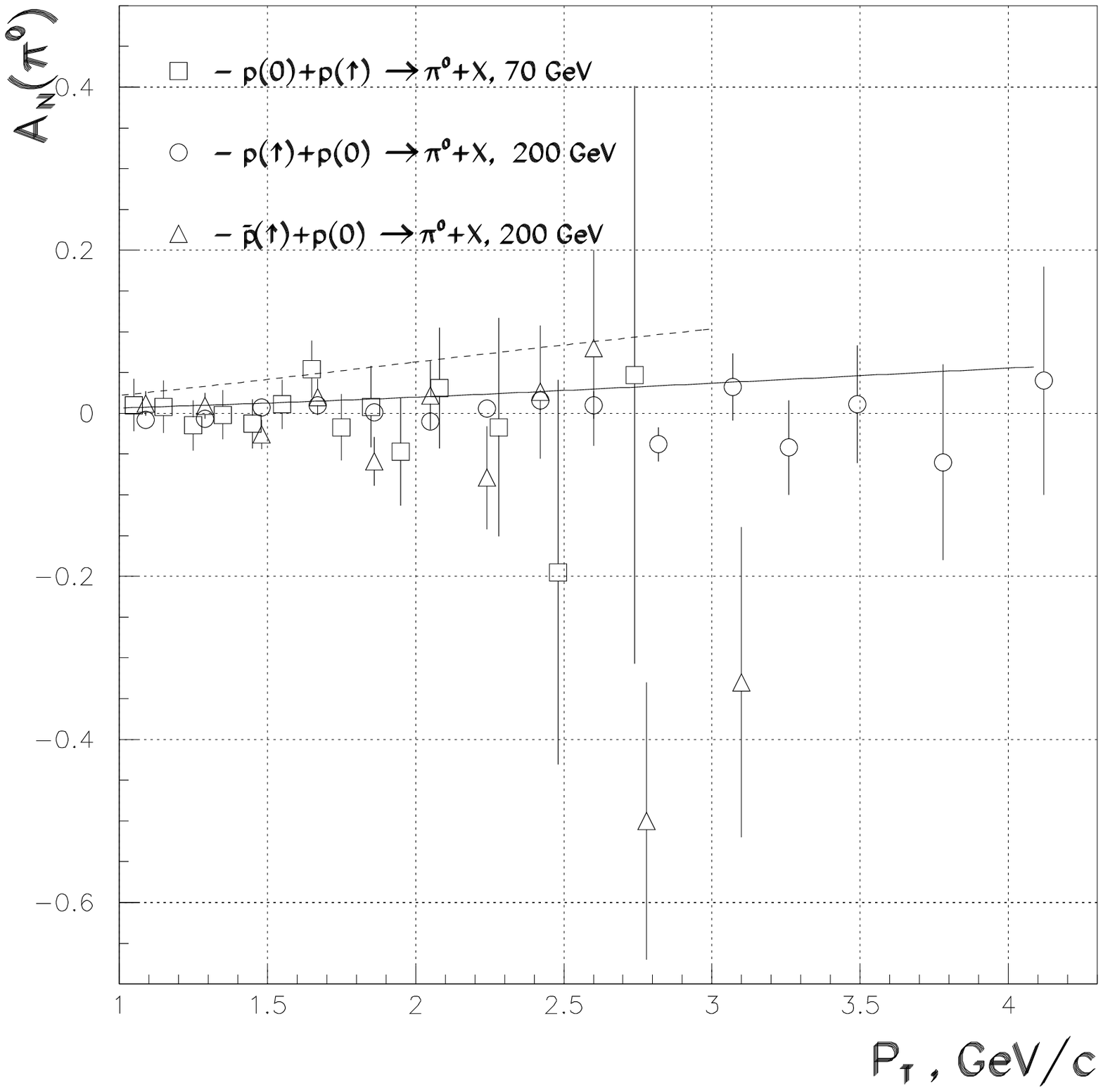} & & 
\includegraphics[width=0.45\textwidth]
{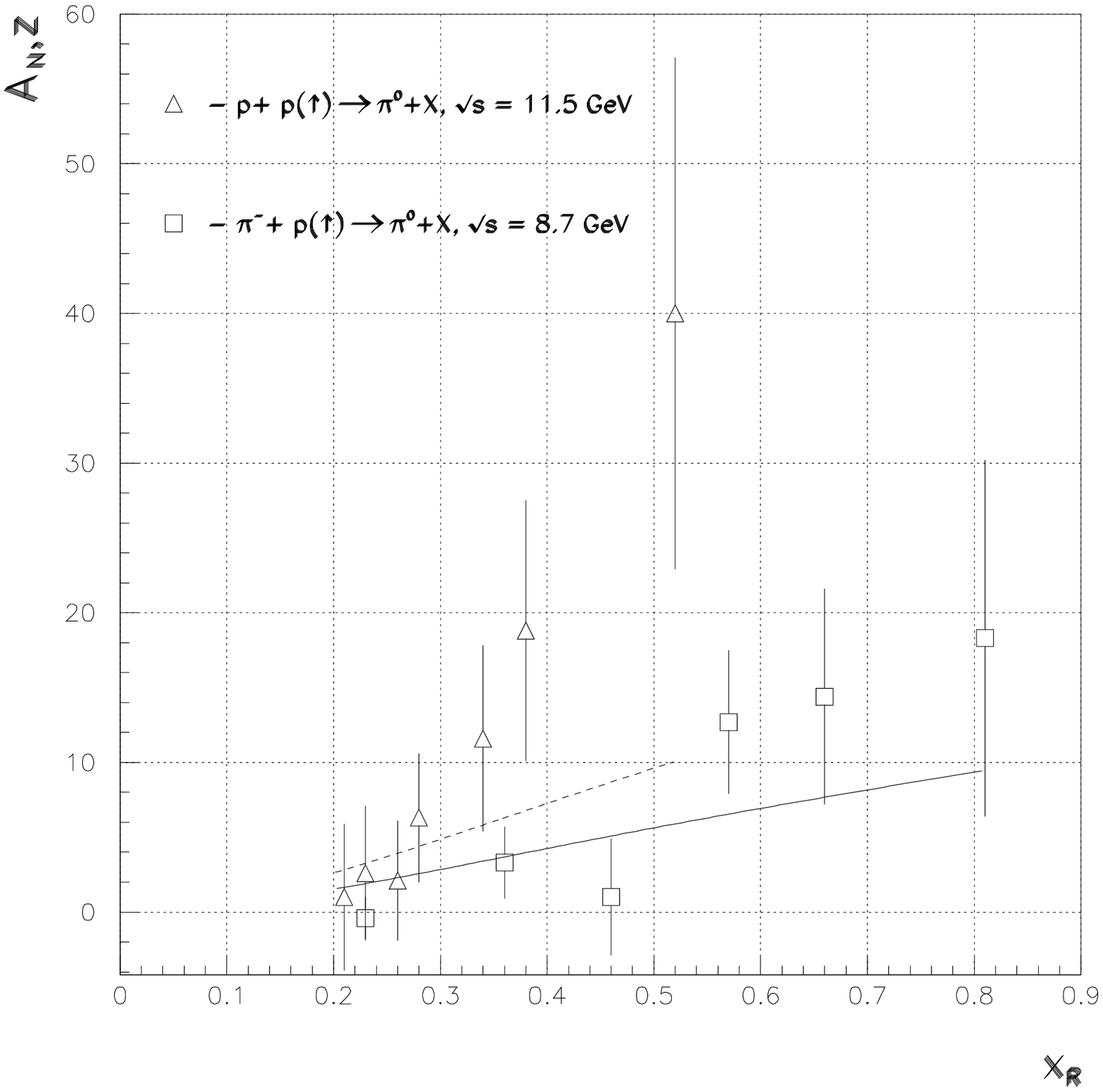}
\end{tabular}
\caption{ SSA vs $p_T$ for reaction \ppdup (left) and SSA vs $x_R$ for reaction  \pimpr(right). Lines are the model prediction (see text) 
\label{asym}}
\end{figure}

New data on  $\pi^-p{\uparrow}\rightarrow \pi^0X$ at
40 GeV/c  are presented in Fig.2b (open square).  
Since in this case $x_F \mbox{ and }x_T$ are comparable in magnitude $A_N(in \%)$
is plotted versus $x_R$ ( $x_R=\sqrt{x_F^2+x_T^2}$) using following formula  
\begin{equation}
A_N(\pi^0)=11.7\frac{x_R^{1.5}}{[x_R^{0.5}+0.64\cdot(1-x_R)^{4.5}]}.
\end{equation} 
 As seen from Fig.2b (solid line) there is
a fairly good description of our data.
In the same Fig.2b the SSA data for reaction $pp_{\uparrow}\rightarrow \pi^0X$ at 70 GeV/c are presented. The detailed discussion was given by Dr. Mochalov in this session
\cite{mochalov}.  The CMSM predicts the same formula as above but  with parameter 20.0 replacing 11.7. In this case the model prediction (the dashed line) does not contradict
to the experimental data in the frame of the given error bars.
\section*{Summary}
The experimental studies of the analyzing powers of 
reactions \ppdup ~at 70 \gev ~and \pimpr ~at 40 \gev
~allow to make the following conclusions.
For the first time the single spin asymmetry was measured in the
polarized target fragmentation region for reaction \pimpr ~at 40 GeV/c.
For kinematical domain  $-0.8<x_F<-0.4,\  1<p_T(GeV/c)<2,\  A_N=-(13.8\pm3.8)\%$, 
while in the region $-0.4<x_F<-0.1,\  0.5<p_T(GeV/c)<1.5$ asymmetry 
is compatible with zero.
Asymmetry measured in the CR is zero for region $1<p_
T(GeV/c)<3$ for reaction \ppdup ~at 70 GeV/c.
The results are consistent with E704 data at 200 GeV/c.

\end{document}